\documentclass[10pt,journal,compsoc]{IEEEtran}

\usepackage[noend]{algpseudocode}
\usepackage{cite}
\usepackage{balance}
\usepackage{siunitx}
\usepackage{graphicx}
\usepackage{epstopdf}
\usepackage{fixltx2e}
\usepackage{commath}
\usepackage{color}
\usepackage{array}
\usepackage{float}
\usepackage{longtable}
\usepackage{algorithm}

\begin{document}

\title{Closed loop application of electroadhesion for increased precision in texture rendering}

\author{Roman~V.~Grigorii,~\IEEEmembership{Member,~IEEE,}
        J.~Edward~Colgate,~\IEEEmembership{Fellow,~IEEE,}
        
\IEEEcompsocitemizethanks{\IEEEcompsocthanksitem The authors are with the Department of Mechanical Engineering in Northwestern University, Evanston, IL 60208. Correspondence e-mail: romangrigorii2015@u.northwestern.edu}
\thanks{Manuscript received September 24, 2019; revised December 10, 2019}}

\markboth{IEEE TRANSACTIONS ON HAPTICS,\hspace{2mm}VOL. 12,\hspace{2mm}NO. 4,\hspace{2mm}OCTOBER-DECEMBER 2019}%
{Shell \MakeLowercase{\textit{et al.}}: IEEE TRANSACTIONS ON HAPTICS}

\IEEEtitleabstractindextext{

\begin{abstract}

Tactile displays based on friction modulation offer wide-bandwidth forces rendered directly on the fingertip. However, due to a number of touch conditions (e.g., normal force, skin hydration) that result in  variations in the friction force and the strength of modulation effect, the precision of the force rendering remains limited. In this paper we demonstrate a closed-loop electroadhesion method for precise playback of friction force profiles on a human finger and we apply this method to the tactile rendering of several textiles encountered in everyday life. 

\end{abstract}

\begin{IEEEkeywords}
surface haptics, electroadhesion, tactile texture, closed loop control
\end{IEEEkeywords}}

\maketitle
\IEEEdisplaynontitleabstractindextext
\IEEEpeerreviewmaketitle
\IEEEraisesectionheading{\section{Introduction}\label{sec:introduction}}

When we interact with objects through touch, a set of wide-band signals originate at the periphery and propagate across skin to be encoded by at least three types of afferents \cite{mcglone2010cutaneous}. Information contained in these signals is unique to the surface texture \cite{romangrigorii2017,manfredi2014natural} and is responsible for its perception \cite{bensmaia2003vibrations}. A remarkable feature of tactile sense is its utilization of both spatially distributed and temporally located features in shaping the percept, which has drawn many comparisons to sight and hearing \cite{saal2016importance}. The well-established link between excitation of the periphery and perceptual consequence that exists for vision and audition would suggest that that capture of skin displacement during interaction with a particular texture and its subsequent playback by means of transducer actuation may be sufficient in informing the brain of the original texture. As it stands today, however, no method of capture and playback encompasses all aspects of touch. This fact becomes starkly apparent when attempting to replicate natural textures. The exact source of the difficulty, whether it is due to fundamental technological limitations in providing sufficiently complex tactile stimulus, controllability of the effect, or the algorithms which command it, remains to be understood.

Several distinct forms of actuation have been developed to offer informative tactile feedback. Array-based tactile displays are one of the earliest and stand out in their unique ability to provide spatially distributed information within the contact patch \cite{velazquez2005low}. Limited in actuator bandwidth and density, this technology is generally best at displaying very coarse textural features. Vibrotactile displays take advantage of the skin's high sensitivity to vibration \cite{verrillo1969sensation} and have been put to good use, primarily for alerts, in many hand held devices. Although vibrotaction is effective in producing highly salient effects, it is typically limited to a narrow range of frequencies and is constrained to transient or vibration type stimulation. Friction modulation displays render tactile features directly on the surface being felt, offering the widest bandwidth and a rich gamut of features including complex textures and shapes. Ultrasonic vibration \cite{biet2007squeeze} and electrostatics \cite{shultz2015surface} have been two approaches for modulation of surface friction and for playback of perceptually salient finger-texture interaction. Accompanied by tribology-based texture capture \cite{romangrigorii2017} friction modulation has great potential to provide a degree of realism in playback of natural texture. 

There are several challenges associated with applying friction modulation in rendering of tactile surface features. On the perceptual end, it fails to render spatially discrete features within the contact patch - a highly informative aspect of touch that SA-I and RA-I afferents are chiefly responsible for encoding. There is strong evidence that this is a minor problem in the case of fine texture rendering and some evidence suggests that coarse spatial features rendered to the whole patch are spatially resolved on the same scale as those resolved within the patch \cite{meyer2015modeling}. Beyond perceptual limitations in rendering capabilities, difficulties exist at the technological end as well. For instance, both effects exhibit nonlinear mappings from electrical signal input to the generated friction force and exhibit variations in the strength of the friction modulation effects due to surface, touch, and skin conditions including sweat secretion, variation of contact area, as well as fingertip surface and bulk material properties. Texture playback by friction modulation, therefore, requires some mitigation of these effects in order to provide good control of the stimulus.

\begin{figure}[t]
\centering
\includegraphics[width=8.5cm, keepaspectratio]{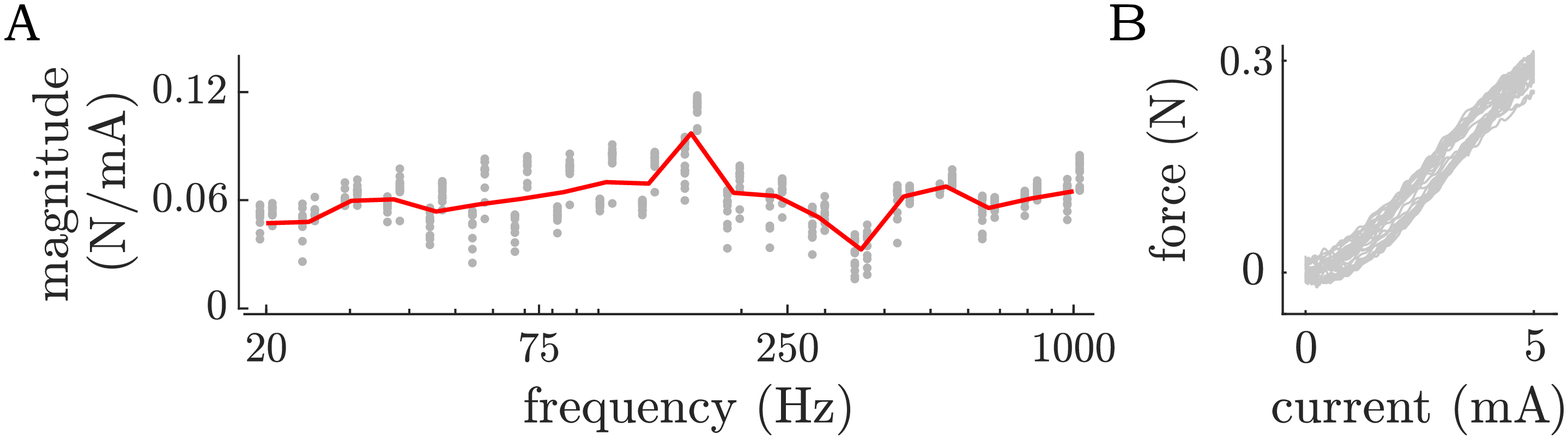}
\caption{A. Additional friction force achieved per mA of applied electroadhesion current across the tactile bandwidth. Grey Points represent mean amplitude across a swipe with data sampled during leftward and rightward swipes shifted respectively about the actuation frequency. Solid red line represents the mean. B. Additional friction force achieved during a single swipe across a 3M surface while a 20kHz electroadhesion current modulated at 10Hz is applied. }
\label{openloop}
\end{figure}

As is the case in any problem that involves tracking performance, one possible solution is to close the loop on the effect itself. This has been realised in the case of friction modulation based on ultrasonic vibration, leading to improved tracking of coarse ($>$32ms, 200mN) surface features  \cite{overcoming2018}, as well in the case of electrovibration for increased uniformity of single frequency actuation \cite{JehaRyu2017}. Electroadhesion has the advantage (compared to ultrasonic vibration) of wide actuation bandwidth without the need for power compensation at higher frequencies and generally offers greater effect stability over the tactile frequency range \cite{Shultz2018}, making it a superior candidate for improving rendering of fine texture signals (on the order of 1ms and 1mN). It is, however, sensitive to touch conditions and is nonlinear as is demonstrated by the results of a pilot experiment in Figure \ref{openloop}, where it is clear
that even swipe direction significantly influences the strength of the effect (possibly due to changes in contact patch area). In the current work we leverage the bandwidth of electroadhesion for improved tracking of the high frequency forces that occur when a fingertip strokes a finely textured surface. 

\section{Methods}
\subsection{Apparatus}
\begin{figure}[t]
\centering
\includegraphics[width = 8cm, keepaspectratio]{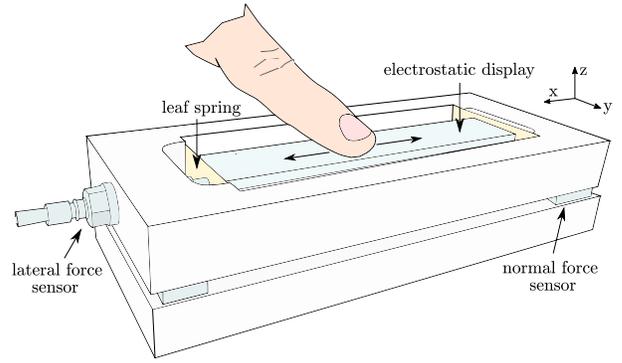}
\caption{Tribometer apparatus.}
\label{apparatus}
\end{figure}

A high-bandwidth tribometer (Figure \ref{apparatus}) was constructed to measure the lateral and normal forces between finger and electrostatic glass (3M Microtouch). The glass, 22 x 105 mm in size, was mounted to an aluminum substrate supported by two leaf springs which were secured inside the milled pocket of an aluminum block - referred to as the housing block. This configuration allowed lateral motion of the surface along its length (x axis) and significantly limited motion in orthogonal directions (y and z axes). The leaf springs pre-loaded the haptic surface against a stiff piezoelectric force sensor (Kistler type 9203) enabling high bandwidth friction force measurements during finger contact. The housing block was positioned on top of four piezoresistive force sensors (Honeywell FSS-SMT) fixed in four corners of a second aluminum block of the same size as the housing block and referred to as the base block. The relative motion of the housing block against the base block was constrained to the normal direction (along z axis) by horizontally placed leaf springs which also uniformly pre-loaded the housing block against the force sensors. The four analog sensor signals were summed into a single value representing the normal load applied by the finger on the haptic surface. Apparatus geometry was specifically designed to eliminate modes of vibration within the tactile frequency range ($<$1kHz) to allow reliable lateral force measurements while minimizing the mass of components coupled to the sensors to maximize sensor bandwidth. The lateral force sensor exhibited a flat frequency response in the DC-1kHz range and the normal force sensor in the DC-250Hz range, with signal linearity of $R^{2} \geq .995$ and $R^{2} > .99$ respectively.

\subsection{Sampling \& computation}

A digital controller was designed and implemented for closed loop control of friction forces generated by electroadhesion effect. A 80MHz microcontroller (PIC32MX) was used for real-time communication and computation of the control signal. Friction force between the finger and the surface and the reference force signal were sampled at 10kHz using a 14bit A/D converter, and the output control signal was generated by 16bit D/A converter. Lateral and normal forces as well as the commanded control signal were sampled at 60kHz by a 16bit DAQ (NU USB-6211) for offline analyses. 

\subsection{Electroadhesion}

Electroadhesion forces were produced by amplitude-modulation of a 20 kHz carrier current.  There are several advantages to this approach as compared to direct modulation of voltages in the tactile frequency range (electrovibration). For instance, at ultrasonic frequencies the gap impedance is largely capacitive; as such, modulation of the input current produces a flat actuation strength across tactile frequency and offers higher stability of the effect strength \cite{Shultz2018}. In addition, frequency doubling in friction force due to carrier rectification occurs outside of tactile frequency range. In this work, the 20kHz carrier was generated and modulated by a function generator (Rigol DG 1032Z), amplified by transconductance current amplifier previously used in \cite{Shultz2018}, and passed to the conductive layer of 3M screen interface to modulate the friction force between the finger and the surface.

\subsection{Analog lateral force filter}

Upon conversion to friction force, the 20kHz electroadhesion current carrier is rectified producing a force component which corresponds to the modulator superimposed with a 40kHz force component along with wide-band harmonics. We seek to control only the friction force within the tactile frequency range (10Hz - 1kHz). Prior to being digitally sampled, this force must be filtered to attenuate the high-frequency force components that are imperceptible but will nonetheless leak onto the lower frequency bands of the digital signal via aliasing. A gross mechanical model of the tribometer apparatus is that of a second-order spring-mass-damper system, which would suggest that it is capable of mechanically attenuating force components of frequencies above the resonant ($\approx$4.4kHz). Indeed, a -40dB attenuation in mechanical admittance from DC was observed at 40kHz coinciding with frequency of the first harmonic of rectified electroadhesion current carrier. The carrier harmonics were therefore significantly attenuated mechanically. Unfortunately, a 40kHz force ripple was still apparent in lateral force measurements, and it was estimated that at least another 35dB attenuation of the carrier ripple would be necessary.  An analog low-pass filter based on the Sallen-Key architecture was designed for this purpose. This filter added 35dB attenuation at 40kHz, which resulted in $\geq$ 75dB attenuation of force components $\geq$ 40kHz while producing minimal phase delay.

\begin{figure}[t]
\centering
\includegraphics[width = 8.5cm, keepaspectratio]{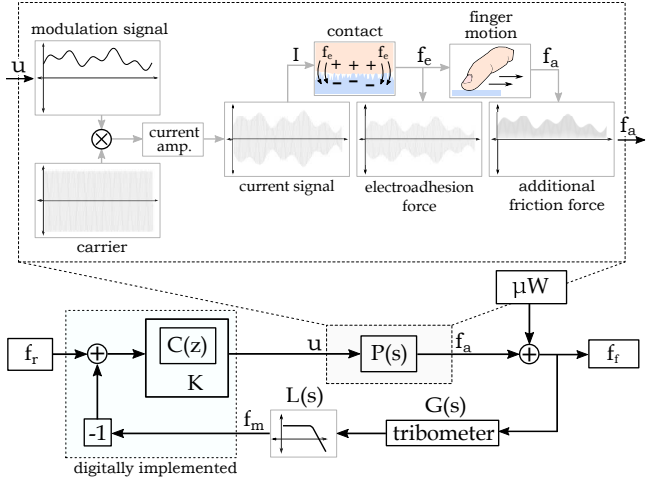}
\caption{Block diagram of the closed loop process. Variables : $\text{f}_{r}$ is the reference force, $\text{f}_{e}$ is electroadhesion force, $\text{f}_{a}$ is additional friction force due to electroadhesion, $\mu$ and $W$ are coefficient of friction and applied normal load respectively, $\text{f}_{f}$ is friction force experienced by the finger, $\text{f}_{m}$ is the measured friction force.}
\label{blockdiagram}
\end{figure}

\section{Closed loop control}

We apply a method of frequency based controller design commonly referred to as loop shaping which offers several advantages in solving the problem of friction force control. First, since the tactile sense is wide-bandwidth in nature, it is beneficial to design a controller based on the desired closed loop frequency response properties. Second, this approach allows us to analytically solve for an optimal controller of a desired order rather than tune controller parameters manually. 

Our goal is to achieve good tracking performance of reference friction force components in the 10Hz - 1kHz range. To this end, we first designed a function, K, to  remove  DC friction during formulation of the tracking error. We then controlled the resulting modified friction force with a digital controller, $C(z)$. The overall block diagram of this process is shown in in Figure \ref{blockdiagram}. To find the optimal controller, the frequency response of all sensors, actuators, and filters in our system were obtained. Digital and analog converters and amplifiers were assumed to have trivial dynamics in the frequency range of interest. We further assumed that all components of the closed loop system were approximately linear time invariant (LTI) which allowed us to approximate their analytic closed form solutions and directly solve for the controller transfer function.  These steps are elaborated in the following sections.

\subsection{System identification}

\begin{figure}[t]
\centering
\includegraphics[width = 8.5cm, keepaspectratio]{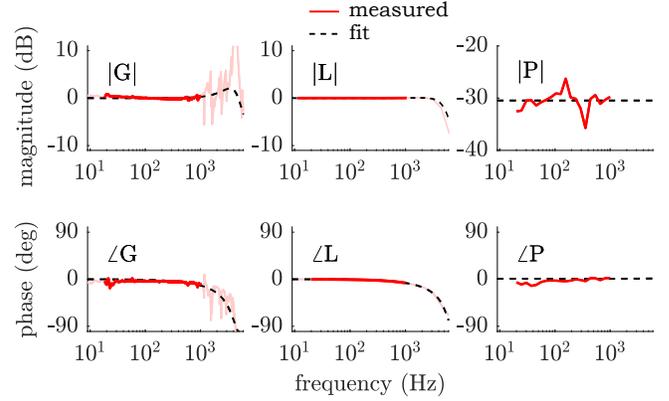}
\caption{Characterized components of the closed loop transfer function. Red solid line represents the measured data and the black dashed line represents an analytical fit.}
\label{characterization_fig}
\end{figure}

To characterize the tribometer, fifteen impulse responses were sampled upon the release of a 2g ball bearing pendulum on the side of the 3M glass. Fourier transforms were computed for each 0.5s long force signal immediately following the impact, with the resulting frequency domain responses indicating the tribometer dynamics. To account for variations in strength of the impact, the magnitude of each frequency response was shifted to match power in the 10Hz-1kHz range (across trials, these shifts varied by no more than 3dB). A second order system was optimally fit to the mean frequency response in the 10Hz - 1kHz range, and the resulting analytic expression is referred to as $G(s)$. 

The analog Sallen-Key low-pass filter was characterized by passing twenty, 2s long, single frequency 1V signals in the 20Hz - 6kHz range (logarithmically spaced). A digital lock-in technique was used to find the magnitude and phase relationship between the input and output signals. The discrete frequency response data was fit by a second order function in the DC-6kHz range, and the analytical solution is referred to as $L(s)$. 

Additional friction achieved by application of electroadhesion force is nonlinear and is a function of touch conditions. Nevertheless, some characterization of the effect in frequency is helpful in controller design. Twenty 10s long, 1.25mA single frequency signals, were used as modulators of the 20kHz carrier and passed through the electroadhesive surface while main author stroked his finger across it. The modulator was biased by 2.5mA, resulting in current signals which ranged between 1.25mA and 3.75mA in magnitude, in order to move actuation into the linear region of the current - force relationship. The frequencies of each signal were logarithmically spaced in the 20Hz - 1kHz range and were pseudo-randomly selected for each trial. We found that the response of output friction force to input current can be approximated by a gain term, $P(s) = .06$ N/mA, as shown in Figure \ref{openloop}. Large variance in $P(s)$ ($P_{min} =.01$ N/mA, $P_{max} =.14$ N/mA) indicates that special care must be taken when formulating the controller in order to keep the system response stable. The frequency responses of the system components and their respective analytic fits are shown in Figure \ref{characterization_fig}.

\begin{algorithm}[t]
\begin{algorithmic}[0]
\small
\vspace{1mm}
\State ${\text{N} \gets 100 \text{ (corresponds to 10ms)}}$
\vspace{2mm}
\State {sample $\text{f}_{m},\text{f}_{r}$}
\vspace{1mm}
\If {patch state $is$ stuck $or$ partially slipping}
    \State $\text{u} \gets 2.5$
    \State $\text{f}^{*} \gets 0$
\Else
    \If {n $\le$ N}
        \vspace{1mm}
        \State $\text{f}^{*} \gets \text{f}^{*} + \text{f}_{m}/{N}$
        \vspace{1mm}
        \State $\text{n} \gets \text{n} + 1$
        \vspace{2mm}
        \State $\text{u} \gets 2.5$
    \Else
        \State $\text{u} \gets C(z)(\text{f}_{r} - \text{f}_{m} + \text{f}^{*})$
        \vspace{2mm}
    \EndIf
\EndIf
\caption{ : operaton of DC friction force mitigator, K}
\end{algorithmic}
\label{algorithm1}
\end{algorithm}

\subsection{DC friction mitigator, K}

Upon change in swipe direction, a finger that is in contact with topographically smooth, flat surface transitions through three contact states: stuck, partially slipping, and fully slipping. These transitions can be located in real time from friction force alone, as previously realized for surface enhancement purposes in \cite{romangrigorii2019}. We adapt the methods of that paper for the purposes of DC friction mitigation. Upon transition of the finger into the full slip state, 10ms of friction force is sampled at 10kHz (equivalent to retaining 100 samples) and is averaged to obtain an estimate of the mean friction force, f*. During this time, the control signal is set to the value that lies half-way along its dynamic range of 2.5mA. At the end of the sampling period the friction controller turned on and is applied to difference of the measured friction force and f* for the duration of the the swipe. When the swipe direction is changed, and the finger patch transitions through the three contact states, f* is recomputed and subtracted from further samples. The pseudo-code for this operation, performed at 10kHz, is presented in Algorithm \ref{algorithm1}.

\subsection{Digital controller, C}

Several considerations must be made during design of the loop shaping controller. First, because the controller is to be implemented in discrete time it will be only an approximation to any continuous time controller that is found directly. Therefore, we utilize an emulation-based design approach in which the controller solution is found by comparison of the discrete time controller sampled at discrete frequencies to the ideal continuous time controller sampled at discrete frequencies. Second, we consider variations in $P(s)$ in order to guarantee stability of the controller during large swings from the expected value of $\overline{P}(s)$ = .06. Third, we keep the order of the controller relatively low due to digital rounding errors that result in large variations in controller response - a situation also prone to instability. 

To solve for the controller, we start with sensitivity function, $T(s)$, of the closed loop system which can be expressed as: 

\begin{equation}
    T = \frac{\mathcal{L}(\text{f}_{\text{f}}(t))}{\mathcal{L}(\text{f}_{\text{r}}(t))}=\frac{{F}_{\text{f}}}{{F}_{\text{r}}}
\end{equation}

Where $F_{r}(\text{s})$ and $F_{\text{f}}(s)$ are Laplace transforms of $\text{f}_{\text{r}}(t)$ and $\text{f}_{\text{f}}(t)$ - the reference force and the resulting friction force respectively. According to the block diagram in Figure \ref{blockdiagram}, $T(s)$ can be written as function of the system components in Laplace domain, namely:

\begin{equation}
    T = \frac{CP}{1+CPLG}
\end{equation}

Setting $T(s)$ = 1, which corresponds to ideal tracking, and solving for $C(s)$ we find:

\begin{equation}
    C = \frac{T}{P(1-TLG)}
\end{equation}

This controller achieves loop shaping, but will not, in general, be stable.  To achieve both stability and reasonable loop shaping, $C(s)$ is modified such that $T(s)$ exhibits the following properties in the 10 $\leq$ $f$ $\leq$ 1000 range:

\begin{enumerate}
\item log$|T(2\pi jf)|$  = $\epsilon(2\pi jf)$
\item arg(${T(2\pi jf)})$ = $\gamma(2\pi jf)$
\end{enumerate}

 Where we seek to minimize deviation of the closed loop transfer function, $T(s)$, from an ideal case where there is no difference in magnitude and phase between resulting friction and the reference, i.e. $\epsilon$ = $\gamma$ = 0. A third order discrete time controller was designed and tuned to match magnitude and phase of the ideal continuous time controller over the widest possible frequency range. To account for uncertainty in $P(s)$, the gain term of the resulting controller was then scaled back by a factor of 2.5 which reduces the bandwidth of the controller but ensures stable tracking. This is equivalent to assuming that $P(s) = P_{max}$. It should be noted that when the effect strength, $P(s)$, is further reduced (e.g. due to reduction of gap impedance with accumulation of moisture on the finger patch), the bandwidth of resulting sensitivity function is also reduced. 

\begin{figure}[t]
\centering
\includegraphics[width = 8.5cm, keepaspectratio]{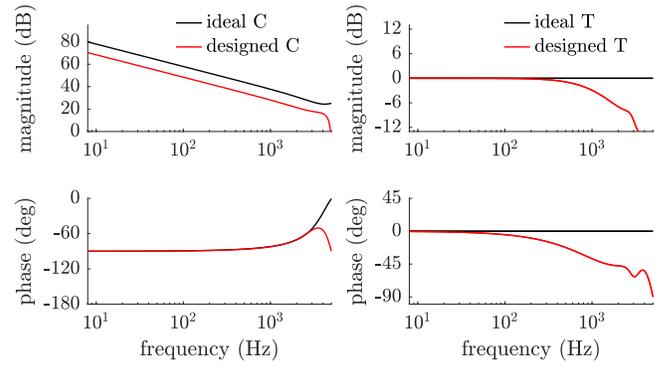}
\caption{Transfer functions of the ideal and designed controller and resulting closed loop response.}
\label{CT}
\end{figure}

A continuous time controller designed to exhibit the same frequency characteristics as the discrete time controller was used to estimate the theoretical performance of $T(s)$. $T(s)$ exhibited a 1kHz, 3dB bandwidth, which is sufficient for fine texture control over the entire tactile bandwidth. The ideal and designed frequency responses are shown in Figure \ref{CT}. 

\subsection{System disturbance \& noise}

One of chief disturbances in the system originates from the native friction of the 3M surface (the average value of which is referred to in Figure \ref{blockdiagram} as $\mu$W, the product of coefficient of friction and applied normal load). Friction forces are noisy, typically exhibiting a spectral magnitude that decreases linearly with increasing frequency on a log scale.  This was confirmed by swiping the lead author's finger across the surface under 0.5N of applied normal load. At 10Hz, a noise level of $10^{-3}$ N was found, and at 1kHz, a noise level of $10^{-5}$ N was found. The noise in the native friction force sets an approximate lower limit of controllable lateral forces across the tactile spectrum using friction modulation. 

Additional noise sources were the drifts of the piezoelectric lateral force sensor and the current amplifier. The signal-to-noise ratio for the 1 sec long texture signals used in this study and the sensor noise was found to be $>$ 30, which was deemed sufficiently accurate for our purposes. Noise propagation analysis demonstrated that the controller did not adversely amplify this noise due its low-frequency nature.

\section{Experimentation and analysis}

Several experiments were conducted to verify the controllability of friction forces applied to the lead author's finger during active touch. To observe the general behavior of the controller, a 20Hz, 25mN square wave was set as the reference force while the lead author swiped his finger back and forth across the screen in a kinematically unconstrained fashion. The resulting friction force was then low-passed with a second order, zero-phase Butterworth filter having a corner frequency of 1kHz.  To verify the effective sensitivity function, $T(s)$, twenty 10s long sinusoids with frequencies ranging logarithmically from 20Hz to 1kHz and four amplitudes (10,20,30, and 40 mN), were rendered to the lead author's finger as he swiped it back and forth across the screen. Individual swipes were distinguished by zero-crossing of friction force, with the middle 50\% of each swipe retained for further analysis. A digital lock-in technique was used to extract friction force amplitude and phase information.

To verify the controller's ability to reliably track fine texture forces, friction force data collected from six natural textures in \cite{romangrigorii2017} were rendered under open and closed loop conditions and then compared. Friction signals collected from each natural texture were "stitched" together using a feathering technique to form 20s long temporal reference signals. The resulting friction force sampled under open and closed loop rendering conditions was band-passed between 10Hz and 1kHz to remove DC force and tribometer dynamics. Friction force data was then shifted forward to match the point of largest correlation with the reference force signal ($\approx.4$ms in time for the closed loop case) for easier evaluation of closed loop tracking performance. This procedure is deemed reasonable in light of the fact that such a small delay would not be tactilely detectable. 
 
\section{Results}

\begin{figure}[t]
\centering
\includegraphics[width = 8.5cm, keepaspectratio]{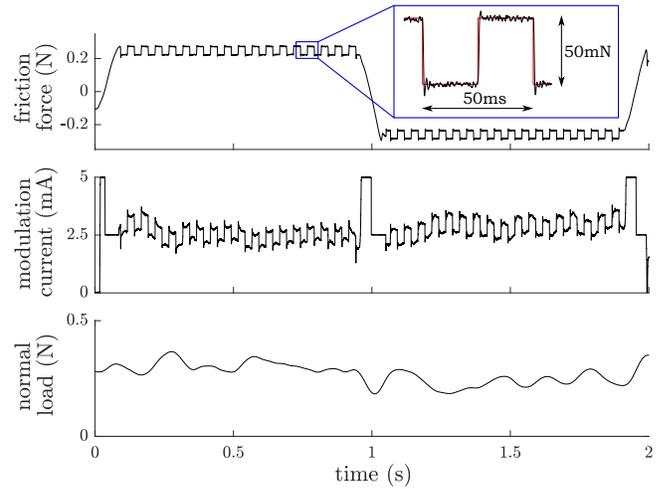}
\caption{a) Friction force (black) and force reference (red), b) control electroadhesion current, and c) applied normal load during a left and right finger swipe while applying closed loop controlled force with a 25mN 20Hz square wave as reference.}
\label{swipe}
\end{figure}

\begin{figure}[b]
\centering
\includegraphics[width = 8.5cm, keepaspectratio]{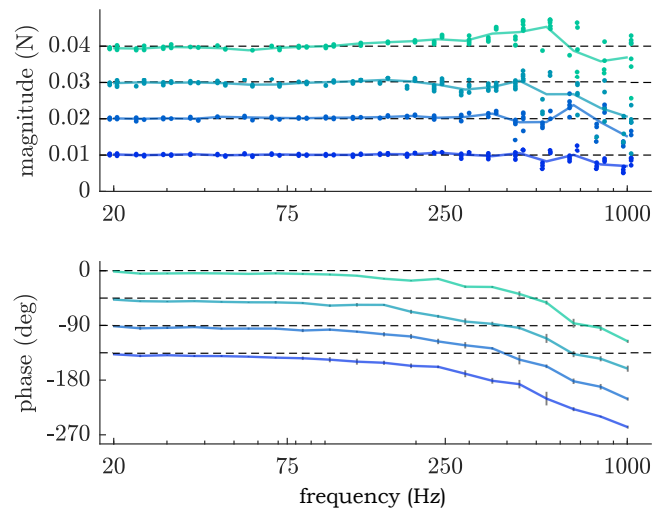}
\caption{Empirically found magnitude and phase response of the closed loop controller. Phase plots for each magnitude were shifted by 45 degrees for easier visualization. Black dashed lines represent ideal controller performance, points represent mean amplitude found during given swipe, solid lines represent mean amplitude across all swipes, and vertical lines on the phase plot represent standard deviation.}
\label{outputtf}
\end{figure}

Figure \ref{swipe} demonstrates the closed loop tracking of a 20Hz 25mN square wave reference. The controller works remarkably well in actuating the fingerpad in a consistent manner across a given swipe, irrespective of swipe direction and applied normal load. It can be observed that upon detection of a stuck state, the controller takes the value of 2.5mA and retains it until an average friction force during full slip state is found, which is then subtracted from all subsequent sampled friction forces. In order to evaluate the sensitivity function, $T(s)$, as it pertains to closed loop control friction force signals in the tactile frequency range, the lead author performed multiple swipes over the surface during closed loop playback of sinusoidal friction forces of different amplitudes and frequencies, with results shown in Figure \ref{outputtf}. We find that there is very good agreement between reference signal and rendered signal for all four magnitudes tested up to 250Hz, after which the controller is unable to keep up with variations in the reference. At or below 100Hz the phase delay does not exceed 1ms, indicating that coarse texture signals will be reliably reproduced in time (or space, under a spatial texture rendering protocol). Some deviation in the phase from the theoretical values (Figure \ref{CT}) exists and may correspond to the delay between reading in the reference signal and updating the force ($\approx .4ms$).

\begin{figure}[t]
\centering
\includegraphics[width = 8.5 cm,keepaspectratio]{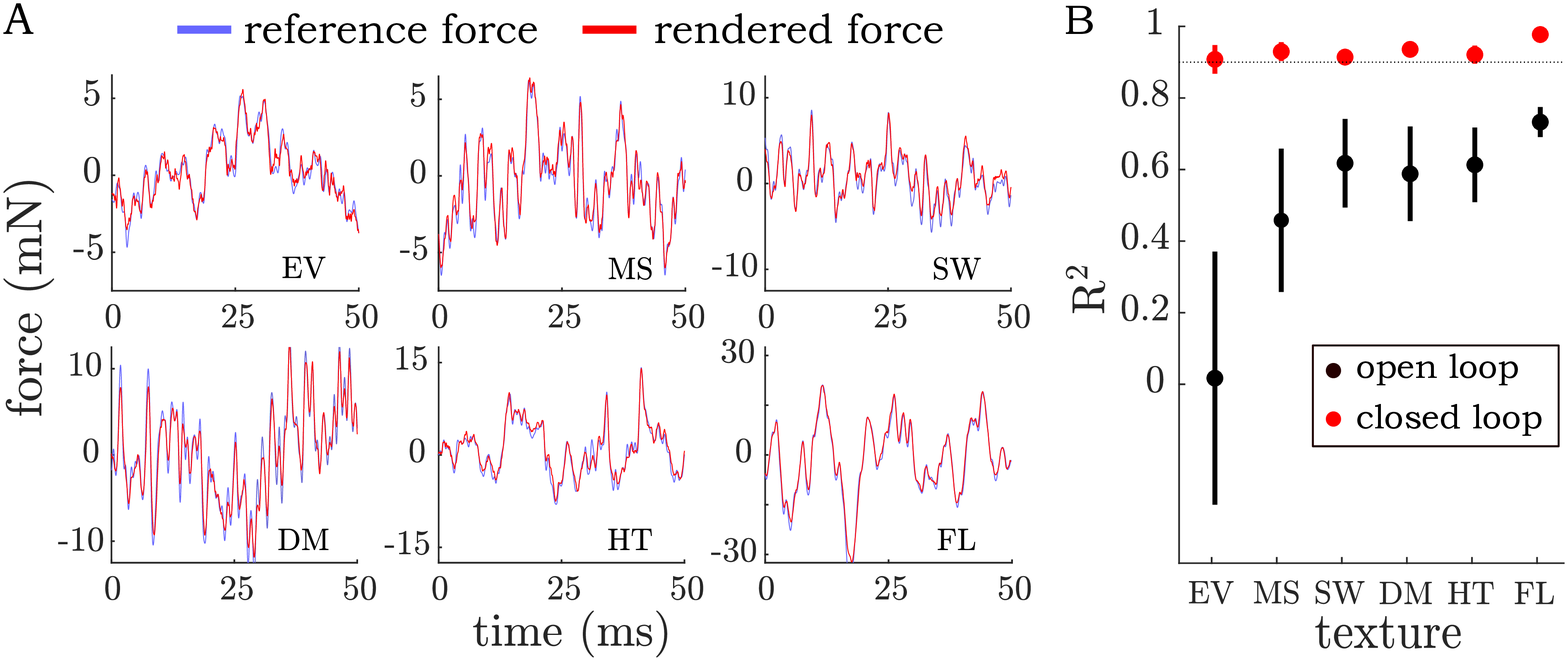}
\caption{A) Closed loop rendering of shear forces elicited by natural textures: Empire Velveteen (EV), Microsuede (MS), Swimwear (SW), Denim (DM), Hucktowel (HT), and Faux Leather (FL) B) Goodness of linear fit between reference force and measured force across textures.}
\label{texturecontrol}
\end{figure}

Closed loop tracking of the six reference signals obtained from natural textures can be observed in Figure \ref{texturecontrol}. The controller reliably compensates for variation in the electroadhesion effect strength and system disturbance to achieve a degree of precision in playback. Closed loop control is able to track $\> 90 \%$ (max $97\%$) of friction forces sampled from the six natural textures. Closed loop control demonstrates improved tracking compared to open loop,  mitigating large disturbances introduced by variation in normal load and variations in $P(s)$. Closed loop rendering appears to perform especially well in tracking very fine texture forces such as those elicited from Micro-suede, where variation in $P(s)$ significantly dominates the variations in reference signal. These results suggest that a number of fine and coarse texture signals can be rendered using this closed-loop technique.

\section{Discussion}

Despite large improvements in reference signal tracking, we found that the rendered force profiles generally did not help the user recognize the texture being conveyed. An exception to this is that overall effect strength proved highly salient, enabling some clear distinctions to be made (e.g. Hucktowel could be easily distinguished from Empire Velveteen). In general, however, a clear sense of realism in playback was missing and this can be explained in several ways. One, it is possible that temporal playback does not offer the same perceptual salience as a spatial playback. Two, a causal relationship between user inputs (normal load and velocity) and the resulting forces can be an important perceptual factor that is not accounted for in the current rendering paradigm.  For instance, the amplitude of skin excitation scales with velocity and normal load during interaction with natural texture and this scaling may contribute to realism in virtual renderings. In order to test the true perceptual consequence of fine friction control in texture rendering, an experiment must be conducted in which user inputs (velocity, normal load, angle of incidence, etc.) are carefully controlled. Nevertheless, we consider the presented method to be a building block.  In the future, it may be applied to more sophisticated rendering paradigms that are consistent with perceptual decoding process, leading to an improved sense of realism in natural texture rendering.

\section{Conclusion}

A method for closed loop playback of measured friction forces has been developed and tested under unconstrained finger kinematics. The proposed method can be utilized for precise playback purposes, including our own work that aims to elucidate those aspects of tactile excitation which are most essential for texture rendering.

\section*{Acknowledgements}
The material is based upon work supported by the National Science Foundation grant number IIS-1518602. Special thanks goes to Craig Shultz for providing current controller used in the study.

\bibliography{bib.bib}
\end{document}